%% file: Parallel_Suffix_Array_Construction_by_Accelerated_Sampling.tex
\documentclass[a4paper,onecolumn]{llncs}

\pdfoutput=1

\usepackage{url}
\usepackage{a4wide}
\usepackage{xfrac}
\usepackage{amsmath}
\usepackage{amsfonts}
\usepackage{wasysym}
\usepackage{stmaryrd}

\newcommand{\ie}{i.e.\ }
\newcommand{\eg}{e.g.\ }
\newcommand{\srange}[2]{\left [ #1 : #2 \right )}
\newcommand{\frange}[2]{\left [ #1 : #2 \right ]}

\input{Title}

\begin{document}
\maketitle

\input{Abstract}
\input{Introduction}

\input{Difference_Covers}

\input{Sequential_Algorithm}

\input{BSP_Model}

\input{BSP_Algorithm}

\input{Conclusion}

\bibliography{bibliography} 
\bibliographystyle{plain}

\end{document}

%% file: Title.tex
\title{Parallel Suffix Array Construction by Accelerated Sampling\thanks{Research supported by the Centre for Discrete Mathematics and its Applications (DIMAP), University of Warwick, EPSRC award EP/D063191/1}}
\author{Matthew Felice Pace \and Alexander Tiskin}
\institute{DIMAP and Department of Computer Science\\
University of Warwick, Coventry, CV4 7AL, UK}

%% file: Abstract.tex
\begin{abstract}
A deterministic BSP algorithm for constructing the suffix array of a given string is presented, based on a technique which we call \textit{accelerated sampling}. It runs in optimal $O(\frac{n}{p})$ local computation and communication, and requires a near optimal $O(\log \log p)$ synchronisation steps. The algorithm provides an improvement over the synchronisation costs of existing algorithms, and reinforces the importance of the sampling technique.
\end{abstract}


%% file: Introduction.tex
\section{Introduction}
\label{sec:Intro}
Suffix arrays are a fundamental data structure in the string processing field. They have been researched extensively since their introduction by Manber and Myers \cite{ManberMyers,PuglisiSmythTurpin}.

\begin{definition}
Given a string $x = x[0] \ldots x[n-1]$ of length $n \geq 1$, defined over an alphabet $\Sigma$, the suffix array problem aims to construct the suffix array $SA_x = SA_x[0] \ldots SA_x[n-1]$ of $x$ which holds the ordering of all the suffixes $s_i = x[i] \ldots x[n-1]$ of $x$ in ascending order; i.e.  $SA_x[j] = i$ iff $s_i$ is the $j^{th}$ suffix of $x$ in ascending lexicographical order.
\end{definition}

\subsection{Notation, Assumptions and Restrictions}

We assume zero-based indexing throughout the paper, and that the set of natural numbers includes zero. For any $i, j \in \mathbb{N}$, we use the notation $\frange{i}{j}$ to denote the set $\{a \in \mathbb{N} \mid i \leq a \leq j\}$, and $\srange{i}{j}$ to denote $\{a \in \mathbb{N} \mid i \leq a < j\}$. 

The input of the algorithms to be presented in this paper is restricted to strings defined over the alphabet $\Sigma = \srange{0}{n}$, where $n$ is the size of the input string. This allows us to use counting sort \cite{CLRS} throughout when sorting characters, in order to keep the running time linear in the size of the input. Counting sort is also used in conjunction with the radix sorting technique \cite{CLRS}.

The set notation described above is extended to substrings by denoting the substrings of string $x$ by $x\srange{i}{j}$, where $x\srange{i}{j} = x[i] \ldots x[j-1]$. Also, the end of any string is assumed to be marked by an end sentinel, typically denoted \$, that precedes all the characters in the alphabet order. Therefore, to mark the end of the string and to ensure that any substring $x\srange{i}{j}$ is well defined, for $i \in \srange{0}{n}$ and $j > i$, we let $x[k] = -1$, for $k \geq n$.

It should be noted that the algorithms to be presented in Sections \ref{sec:SeqAlg}, \ref{sec:ParAlg} can also be applied to any string $X$, of size $n$, over an indexed alphabet $\Sigma'$ \cite{PuglisiSmythTurpin,Smyth}, which is defined as follows:

\begin{itemize}
	\item $\Sigma'$ is a totally ordered set.
	\item an array $A$ can be defined, such that, $\forall \sigma \in \Sigma'$, $A[\sigma]$ can be accessed in constant time.
	\item $|\Sigma'| \leq n$.
\end{itemize}

Commonly used indexed alphabets include the ASCII alphabet and the DNA bases. It should also be noted that any string $X$, of size $n$, over a totally ordered alphabet can be encoded as a string over integers. This is achieved by sorting the characters of the string, removing any duplicates, and assigning a rank to each character. A new string $X'$ of size $n$ is then constructed, such that it is identical to $X$ except that each character of $X$ is replaced by its rank in the sorted list of characters. However, sorting the characters of $X$ could require $O(n \log n)$ time, depending on the nature of the alphabet over which $X$ is defined.

The example in Table \ref{tbl:example} shows the suffix array for a string $X$, of size 12, over an indexed alphabet of a subset of the ASCII characters, written as string $X'$ over $\Sigma = \srange{0}{12}$.

\begin{table}[h]
\centering
\caption{Suffix array of a string $X$ over an indexed alphabet, written as $X'$ over $\Sigma = \srange{0}{12}$}
\begin{tabular}{rp{5mm}p{5mm}p{5mm}p{5mm}p{5mm}p{5mm}p{5mm}p{5mm}p{5mm}p{5mm}p{5mm}p{5mm}p{5mm}}
&$0$&$1$&$2$&$3$&$4$&$5$&$6$&$7$&$8$&$9$&$10$&$11$&$12$\\
$X=$&$a$&$c$&$b$&$a$&$a$&$c$&$e$&$d$&$b$&$b$&$e$&$a$&$\$$\\
$X'=$&$0$&$2$&$1$&$0$&$0$&$2$&$4$&$3$&$1$&$1$&$4$&$0$&$-1$\\
$SA_X=$&$11$&$3$&$0$&$4$&$2$&$8$&$9$&$1$&$5$&$7$&$10$&$6$\\
\end{tabular}
\label{tbl:example}
\end{table}

Let $x_1 \odot x_2$ denote the concatenation of strings $x_1$ and $x_2$. Then, for any set of integers $A$, $\bigodot_{i \in A} x_i$ is the concatenation of the strings indexed by the elements of $A$, in ascending index order. Throughout the paper we use $|b|$ to denote the size of an array or string $b$. To omit $\lceil \cdot \rceil$ operations, we assume that any real numbers are rounded up to the nearest integer.

\subsection{Problem Overview}

The suffix array problem is, by definition, directly related to the sorting problem. In fact, if all the characters of the input string are distinct, then the suffix array is obtained by sorting the strings' characters and returning the indices of the characters in their sorted order. In general, if the characters of the string are not distinct, the naive solution is to radix sort all the suffixes, which takes $O(n^2)$ time if counting sort is used to sort the characters at each level of the radix sort. However, numerous algorithms exist that improve on this running time. The first such algorithm was presented by Manber and Myers \cite{ManberMyers} and required $O(n \log n)$ time. The running time was reduced to $O(n)$ through three separate algorithms presented by K\"{a}rkk\"{a}inen and Sanders \cite{KarkkainenSanders}, Kim et al. \cite{KimSimParkPark}, and Ko and Aluru \cite{KoAluru}. A number of other algorithms exist with a higher theoretical worst case running time but faster running time in practice, as discussed in \cite{PuglisiSmythTurpin}. However, the study of these is beyond the scope of this work.

The idea behind the algorithms having linear theoretical worst case running time is to use recursion as follows:

\begin{enumerate}
	\item Divide the indices of the input string $x$ into two nonempty disjoint sets. Form strings $x'$ and $y'$ from the characters indexed by the elements of each set. Recursively construct $SA_{x'}$.
	\item Use $SA_{x'}$ to construct $SA_{y'}$.
	\item Merge $SA_{x'}$ and $SA_{y'}$ to obtain $SA_x$.
\end{enumerate}

The problem of constructing suffix arrays, while similar to the sorting problem, differs as follows. Given two sorted lists of integers, we are guaranteed that after merging them the order of the integers in the original lists is preserved. However, given two strings and their suffix arrays, the order of the suffixes is not necessarily preserved in the suffix array of the concatenated string. For example, the suffix arrays of strings $aaa$ and $aab$ are $[2,1,0]$ and $[0,1,2]$ respectively, but the suffix array of string $aaaaab$ is $[0,1,2,3,4,5]$.

The aim of this paper is to investigate the suffix array construction problem in the Bulk Synchronous Parallel (BSP) model, on a $p$ processor distributed memory system. As in the sequential setting, the naive general solution to the problem is to radix sort all the suffixes of the string. Shi and Shaeffer \cite{ShiSchaeffer} provide a comparison based parallel sorting algorithm, using a technique known as regular sampling, which is then adapted by Chan and Dehne \cite{ChanDehne} for integer sorting. However, using such a technique to sort the suffixes of a given string of size $n$ leads to a parallel algorithm with $O(\frac{n^2}{p})$ local computation cost, $O(n)$ communication cost and requiring $O(1)$ synchronisation steps. Clearly, it is more efficient to simply use a linear time sequential algorithm.

K\"{a}rkk\"{a}inen et al. \cite{KarkkainenSandersBurkhardt} give a brief overview of a BSP suffix array construction algorithm having optimal $O(\frac{n}{p})$ local computation and communication costs and requiring $O(\log^2 p)$ synchronisation steps. They also present similar algorithms in the PRAM model. In this paper we further reduce the number of synchronisation steps required to a near optimal $O(\log \log p)$, while keeping the same optimal local computation and communication costs. The algorithm is based on a technique that we call \emph{accelerated sampling}. This technique was introduced (without a name) by Tiskin \cite{Tiskin} for the parallel selection problem. An accelerated sampling algorithm is a recursive algorithm that samples the data at each level of recursion, changing the sampling frequency at a carefully chosen rate as the algorithm progresses.

\subsection{Paper Structure}

The rest of the paper is structured as follows. The next section provides an overview of the concept of difference covers. The sequential suffix array construction algorithm is given in Section \ref{sec:SeqAlg}. An overview of the BSP model is provided in Section \ref{sec:BSP}, and a description of the parallel suffix array construction algorithm in this model is presented in Section \ref{sec:ParAlg}. The last section offers some concluding views and discusses possible future work.

%% file: Difference_Covers.tex
\section{Difference Covers}
\label{sec:DiffCov}
The suffix array construction algorithms to be presented in this paper make use of the concept of difference covers \cite{ColbournLing,KilianKipnis,MereghettiPalano}. Given a positive integer $v$, let $\mathbb{Z}_v$ denote the set of integers $\srange{0}{v}$. A set $D \subseteq \mathbb{Z}_v$ can be defined such that for any $z \in \mathbb{Z}_v$, there exist $a, b \in D$ such that $z \equiv a - b$ $(\bmod$ $v)$. Such a set $D$ is known as a \textit{difference cover} of $\mathbb{Z}_v$, or \textit{difference cover modulo $v$} of $\mathbb{Z}_v$.

Colbourn and Ling \cite{ColbournLing} present a method for obtaining, for any $v$, a difference cover $D$ of $\mathbb{Z}_v$ in time $O(\sqrt{v})$, where $|D| = 6r + 4$, $r = \frac{-36 + \sqrt{48 + 96v}}{48}$. Hence, $|D| \leq \sqrt{1.5v} + 6$. Note that, in general, for any $v$ and any difference cover $D$ of $\mathbb{Z}_v$, $|D| \geq \frac{1 + \sqrt{4v - 3}}{2}$, since we must have $|D| (|D| - 1) + 1 \geq v$. Therefore, the size of the difference cover obtained by using the algorithm in \cite{ColbournLing} is optimal up to a multiplicative constant.

The algorithms to be presented in this paper require that $|D| < v$, so we assume $v \geq 3$. The optimal difference covers of $\mathbb{Z}_3$, $\mathbb{Z}_4$ are of size 2, 3 respectively, and for $v \geq 5$ the method of \cite{ColbournLing} gives difference covers of sizes given in Table \ref{tab:DiffCov}.

For technical reasons, discussed in Section \ref{sec:SeqAlg}, we also require that $0 \not \in D$. This does not represent a restriction since, for any $v$ and difference cover $D$ of $\mathbb{Z}_v$, a fixed $z \in \mathbb{Z}_v$ can always be chosen such that the set $D' = \{(d - z) \bmod v \mid d \in D\}$ is also a difference cover of $\mathbb{Z}_v$ (see \eg \cite{MereghettiPalano}).

The following lemma is also required to ensure the correctness of the algorithms to be presented.

\begin{table}[t]
\centering
\caption{Size of the difference cover obtained using the algorithm in \cite{ColbournLing} for various values of $v$}
\begin{tabular}{|c|c|c|c|c|c|c|c|}
\hline
$v$&$5 \ldots 13$&$14 \ldots 73$&$74 \ldots 181$&$182 \ldots 337$&$338 \ldots 541$&$1024$&$2048$\\
\hline
$|D_v|$&$4$&$10$&$16$&$22$&$28$&$40$&$58$\\
\hline
\end{tabular}
\label{tab:DiffCov}
\end{table}

\begin{lemma}\textup{\textbf{\cite{KarkkainenSandersBurkhardt}}}
If $D$ is a difference cover of $\mathbb{Z}_v$, and i and j are integers, then there exists $l \in \srange{0}{v}$ such that $(i + l) \bmod v$ and $(j + l) \bmod v$ are both in $D$.
\label{lem:DiffCov}
\end{lemma}

For any difference cover $D$ of $\mathbb{Z}_v$ and integer $n \geq v$, a \textit{difference cover sample} is defined as $C = \{i \in \srange{0}{n} \mid i \bmod v \in D\}$. The index set $C$ is a $v$-periodic sample of $\srange{0}{n}$, as defined in \cite{KarkkainenSandersBurkhardt}. The fact that difference cover samples are periodic allows them to be used for efficient suffix sorting on a given string.

%% file: Sequential_Algorithm.tex
\section{Sequential Algorithm}
\label{sec:SeqAlg}
K\"{a}rkk\"{a}inen et al. \cite{KarkkainenSandersBurkhardt} present a sequential recursive algorithm that constructs the suffix array of a given string $x$ of size $n$, using a difference cover $D$ of $\mathbb{Z}_v$, for any arbitrary choice of $v \in \frange{3}{n}$, in time $O(vn)$. Clearly, by setting $v = 3$ the running time of the algorithm is $O(n)$, with a small multiplicative constant. As $v$ approaches $n$ the running time approaches $O(n^2)$, and when $v = n$ the algorithm is simply a complex version of the naive suffix array construction algorithm. However, by initially letting $v = 3$ and increasing the value of $v$ at a carefully chosen rate in every subsequent level of recursion, we can reduce the total number of recursion levels required for the algorithm to terminate, while still keeping the total running time linear in the size of the input string. This technique can be used to decrease the number of synchronisation steps required by the parallel suffix array construction algorithm in the BSP model. This is discussed further in Section \ref{sec:ParAlg}. The detailed sequential algorithm proceeds as follows:

\noindent \textit{Recursion base}

We sort $x$ using counting sort, in time $O(n)$. If all the characters of $x$ are distinct we return, for each character, in the sorted order, the index of the character in $x$, \ie $SA_x$. Otherwise, the following steps are performed:\vspace{4mm}

\noindent \textbf{Algorithm 1.} \textit{Sequential Suffix Array Construction}

\noindent \textbf{Parameters:} integer $n$; integer $v \in \frange{3}{n}$

\noindent \textbf{Input:} string $x = x[0] \ldots x[n-1]$ over alphabet $\Sigma = \srange{0}{n}$

\noindent \textbf{Output:} suffix array $SA_x = SA_x[0] \ldots SA_x[n-1]$

\noindent \textbf{Description:}

\noindent \textit{Step 0 - Sample construction and initialisation}

Construct the difference cover $D$ of $\mathbb{Z}_v$ as discussed in Section \ref{sec:DiffCov}. Then, for each $k \in \srange{0}{v}$, define the set $B_k = \{i \in \srange{0}{n} \mid i \bmod v = k\}$. This partitions the set of indices of $x$ into $v$ sets of size about $\frac{n}{v}$. The difference cover sample $C = \bigcup_{k \in D} B_k$ is then constructed. For $i \in C$, we call the characters $x[i]$ \emph{sample characters} and the suffixes $s_i$ \emph{sample suffixes}. We also denote by $S_k$, $k \in \srange{0}{v}$, the set of suffixes $s_i$, $i \in B_k$.

Furthermore, an array $rank$ of size $n+v$ is declared and initialised by $rank[0] = \ldots = rank[n+v-1] = -1$. This array will be used to store the rank of the sample characters of $x$ in the suffix array returned by the recursive call made later in step 1. Only $|C|$ elements of $rank$ will be used, and in fact a smaller array can be used to hold these values. However, we use a larger array to avoid complex indexing schemes relating elements in $rank$ to characters in $x$.

\noindent \textit{Step 1 - Sort sample suffixes}

Let $\overline{\Sigma}$ be an alphabet of super-characters, which are defined to be in 1-1 correspondence with the distinct substrings of $x$ of length $v$: super-character $\overline{x\srange{i}{i+v}}$ corresponds to the substring $x\srange{i}{i+v}$, for all $i \in C$. Therefore, $\overline{\Sigma} \subseteq (\Sigma \cup \{-1\})^v$. Recall from Section \ref{sec:Intro} that, due to the padding convention, any substring $x\srange{i}{j}$ is well-defined, for $i \in \srange{0}{n}$ and $j > i$, and therefore any super-character $\overline{x\srange{i}{j}}$ is also well-defined.

For each $k \in D$, we now define a string of super-characters $X_k$ over $\overline{\Sigma}$, where $X_k = \bigodot_{i \in B_k} \overline{x\srange{i}{i+v}}$ and $|X_k| = \frac{n}{v}$. Then, we construct the string of super-characters $X = \bigodot_{k \in D} X_k$, with $|X| = |D|\frac{n}{v}$. Note that for each $k$, the suffixes of $X_k$ correspond to the set of suffixes $S_k$. The last super-character of $X_k$ ends with one or more $-1$ elements, since $0$ is not allowed to be in the difference cover. Therefore, each suffix of $X$ corresponds to a different sample suffix of $x$, followed by one or more $-1$ characters followed by other characters that do not affect the lexicographic order of the suffixes of $X$. Note that, if $0$ was allowed in the difference cover and $n$ was a multiple of $v$, then the last super-character of $X_k$ would not end with $-1$. 

Recall from Section \ref{sec:Intro} that since the input to the algorithm is a string over natural numbers, the string of super-characters $X$ can be encoded as string $X'$ over $\Sigma' = \srange{0}{|X|}$ using radix sorting, in time $O(v|X|)$, where $|X'| = |X| = |D|\frac{n}{v}$. The order of the suffixes of $X$ can then be found by constructing the suffix array of $X$ by recursively calling the algorithm on the string $X'$ over $\Sigma'$, with parameters $|X'|$ and $v'$, where $v'$ can be chosen arbitrarily from the range $\frange{3}{\min \left ( \frac{v^2}{|D|} - 1, |X'| \right )}$. Thus, $v'$ becomes the value of $v$ in the subsequent recursion level. The bound $v' < \frac{v^2}{|D|}$ ensures that the total work performed by the algorithm is still linear in $n$.

Recall from Section \ref{sec:DiffCov} that we require $|D| < v$. This ensures that $|X| < n$, so the algorithm is guaranteed to terminate, since each recursive call is always made on a shorter string. In fact, if the parameter $v$ remains constant over all the levels of the recursion (say $v = 3$), then in each level the size of the string is reduced by a factor of $\frac{|D|}{v}$ (a factor of $\frac{2}{3}$ for $v = 3$, $|D| = 2$). However, by carefully increasing the value $v$ in every round, within the bounds specified above, we can reduce the number of recursion levels of the algorithm by accelerating the rate of string size reduction in each successive level of recursion, as discussed in detail in Section \ref{sec:ParAlg}.

When the recursive call returns with $SA_{X'}$, this holds the ordering of all the suffixes of $X'$, \ie the ordering of the sample suffixes of $x$ within the set of sample suffixes. Then, for $i \in C$, the rank of $s_i$ in $SA_{X'}$ is recorded in $rank[i]$. Note that the order of the sample suffixes within each set $S_k$, $k \in D$, can be found from $SA_{X'}$.

The total cost of this step is dominated by the radix sorting procedure required to encode string $X$ into $X'$ over  $\Sigma' = \srange{0}{|X|}$, which runs in time $O(|D|n)$.

Note that we can now compare any pair of suffixes by the result of Lemma \ref{lem:DiffCov}. However, this is not sufficient to sort the suffixes of $x$ in linear time, since a different value of $l$ would have to be found for each pair of suffixes and linear time sorting would not be possible. Instead, we perform the following steps.

\noindent \textit{Step 2 - Find the order of the non-sample suffixes within each set $S_k$, $k \in \mathbb{Z}_v \setminus D$}

For each $k \in \mathbb{Z}_v \setminus D$, consider any $l_k \in \srange{1}{v}$ such that $(k + l_k) \bmod v \in D$. For every character $x[i]$, $i \in \srange{0}{n} \setminus C$, define the tuple $t_i = (x[i], x[i+1], \ldots, x[i+l_k -1], rank[i+l_k])$, where $k = i \mod v$. Note that $rank[i +l_k]$ is defined for each $i$, since $rank[a]$, for all $a \in C$, has been found in the previous step and $rank[a] = -1$ for all $a \geq n$.

Then, for each set $B_k$, $k \in \mathbb{Z}_v \setminus D$, construct the sequence of tuples $(t_i)_{i \in B_k}$. Each of the $v - |D|$ constructed sequences has about $\frac{n}{v}$ tuples, with each tuple having less than $v$ elements. The order of the suffixes within $S_k$ is then obtained by independently sorting every sequence of tuples $(t_i)_{i \in B_k}$, using radix sorting.

The total computation cost of this step is dominated by the cost of radix sorting all the sequences, \ie $O \left ( \left (v - |D| \right )n \right ) = O(vn)$.

\noindent \textit{Step 3 - Sort all suffixes by first $v$ characters}

Note that in the previous steps the order of every suffix within each set $S_k$, $k \in \srange{0}{v}$, has been found. Now, let $S^{\alpha}$ be the set of suffixes starting with $\alpha$, for $\alpha \in (\Sigma \cup \{-1\})^v$. Then, every set $S^{\alpha}$ is composed of ordered subsets $S^{\alpha}_{k}$, where $S^{\alpha}_{k} = S^{\alpha} \bigcap S_k$.

All the suffixes $s_i$, $i \in \srange{0}{n}$, are partitioned into the sets $S^\alpha$ by representing each suffix by the substring $x\srange{i}{i+v}$, and sorting these substrings using radix sorting in time $O(vn)$.

\noindent \textit{Step 4 - Merge and complete the suffix ordering}

For all $\alpha \in \Sigma^v$, the total order within set $S^\alpha$ can be obtained by merging the subsets $S^{\alpha}_{k}$, $k \in \mathbb{Z}_v$. This comparison-based $v$-way merging step uses the fact that all the suffixes in $x^\alpha$ start with the same substring $\alpha$, in conjunction with Lemma 1. Due to this lemma, a value $l \in \srange{0}{v}$ exists such that for any $i, j$ the comparison of suffixes $s_i$, $s_j$ only requires the comparison of $rank[i + l]$ and $rank[j + l]$. Having already partitioned the suffixes into sets $S^\alpha$ and found the order of the suffixes within each set $S_k$, $k \in [0,v)$, the suffix array can be fully constructed through this merging process in time $O(vn)$. \Square

All the steps of the algorithm can be completed in time $O(vn)$, and the recursive call is made on a string of size at most $\frac{4}{5}n$, which corresponds to $|D| = 4$, $v = 5$. This leads to an overall running time of $O(vn)$.

%% file: BSP_Model.tex
\section{BSP model}
\label{sec:BSP}
The \emph{bulk-synchronous parallel} (BSP) computation model \cite{Valiant,McColl} was introduced by Valiant in 1990, and has been widely studied ever since. The model was introduced with the aim of bridging the gap between the hardware development of parallel systems and the design of algorithms on such systems, by separating the system processors from the communication network. Crucially, it treats the underlying communication medium as a fully abstract communication network providing point-to-point communication in a strictly synchronous fashion. This allows the model to be architecture independent, promoting the design of scalable and portable parallel algorithms, while also allowing for simplified algorithm cost analysis based on a limited number of parameters.

A BSP machine consists of $p$ processors, each with its local primary and secondary memory, connected together through a communication network that allows for point-to-point communication and is equipped with an efficient barrier synchronisation mechanism. It is assumed that the processors are homogeneous and can perform an elementary operation per unit time. The communication network is able to send and receive a word of data to and from every processor in $g$ time units, i.e. $g$ is the inverse bandwidth of the network. Finally, the machine allows the processors to be synchronised every $l$ time units. The machine is, therefore, fully specified using only parameters $p$, $g$, $l$, and is denoted by $\emph{BSP}(p,g,l)$.

An algorithm in the BSP model consists of a series of \emph{supersteps}, or \emph{synchronisation steps}. In a single superstep, each processor performs a number of, possibly overlapping, computation and communication steps in an asynchronous fashion. However, a processor is only allowed to perform operations on data that was available to it at the start of the superstep. Therefore, in a single superstep, a processor can send and receive any amount of data, however, any received data can only be operated on in the following superstep. At the end of a superstep, barrier synchronisation is used to ensure that each processor is finished with all of its computation and data transfer.

The cost of a BSP superstep on a $\emph{BSP}(p,g,l)$ machine can be computed as follows. Let ${work}_i$ be the number of elementary operations performed by processor $P_i$, $i \in \srange{0}{p}$, in this superstep. Then, the \emph{local computation cost} $w$ of this superstep is given by $w = \max_{i \in \srange{0}{p}}({work}_i)$. Let $h^{out}_i$ and $h^{in}_i$ be the maximum number of data units sent and received, respectively, by processor $P_i$, $i \in \srange{0}{p}$, in this superstep. Then, the \emph{communication cost} $h$ of this superstep is defined as $h = \max_{i \in \srange{0}{p}}(h^{out}_i) + \max_{i \in \srange{0}{p}}(h^{in}_i)$. Therefore, the total cost of the superstep is $w + h \cdot g + l$. The total cost of a BSP algorithm with $S$ supersteps, with local computation costs $w_s$ and communication costs $h_s$, $s \in \srange{0}{S}$, is $W + H \cdot g + S \cdot l$, where $W = \sum_{s=0}^{S-1} w_s$ is the total local computation cost and $H = \sum_{s=0}^{S-1} h_s$ is the total communication cost.

The main principle of efficient BSP algorithm design is the minimisation of the algorithm's parameters $W$, $H$, and $S$. These values typically depend on the number of processors $p$ and the problem size.

%% file: BSP_Algorithm.tex
\section{BSP Algorithm}
\label{sec:ParAlg}
Along with the sequential suffix array construction algorithm, described in Section \ref{sec:SeqAlg}, K\"{a}rkk\"{a}inen et al. \cite{KarkkainenSandersBurkhardt} discuss the design of the algorithm on various computation models, including the BSP model. They give a brief overview of a parallel suffix array construction algorithm, running on a $\emph{BSP}(p,g,l)$ machine, with optimal $O(\frac{n}{p})$ local computation and communication costs and requiring $O(\log^2 p)$ synchronisation steps. The algorithm uses a number of existing parallel sorting and merging algorithms to achieve this result. In the first part of this section we present a deterministic BSP algorithm that preserves these optimal local computation and communication costs while reducing the number of required synchronisation steps to a near optimal $O(\log \log p)$. Following this, a detailed algorithm analysis is presented.

The algorithm described in Section \ref{sec:SeqAlg} initially solves the suffix array construction problem on a sample of the suffixes of the input string, in order to gain important information that is then used to efficiently sort all the  suffixes. Sampling techniques are widely used in various fields ranging from statistics to engineering to computer science. In fact, a number of parallel algorithms exist that use sampling to efficiently solve problems, including sorting \cite{ShiSchaeffer,ChanDehne} and convex hull \cite{Tiskin_convexhull} algorithms. In \cite{Tiskin}, Tiskin presents a BSP algorithm for the selection problem, in which, not only is the data sampled, but, the sampling rate is increased at a carefully chosen rate in successive levels of recursion. This reduces the number of synchronisation steps required by the parallel selection algorithm from the previous upper bound of $O(\log p)$ to a near optimal $O(\log \log p$), while keeping the local computation and communication costs optimal. We make use of this technique, which we call \emph{accelerated sampling}, to achieve the same synchronisation costs for our parallel suffix array construction algorithm, while, again, keeping the local computation and communication costs optimal. In contrast with \cite{Tiskin}, in our algorithm the sampling frequency has to be decreased, rather than increased, in successive levels of recursion.

The algorithms presented in this section are designed to run on a $BSP(p,g,l)$ machine. We denote the sub-array of an array $a$ assigned to processor $\pi \in \srange{0}{p}$ by $a_{\pi}$ and extend this notation to sets, \ie we denote by $A_{\pi}$ the subset of a set $A$ assigned to processor ${\pi}$.

Before detailing our algorithm, we give an overview of the parallel integer sorting algorithm introduced in \cite{ChanDehne}. The algorithm is based on the parallel sorting by regular sampling algorithm \cite{ShiSchaeffer}, but uses radix sorting to locally sort the input, removing the extra cost associated with comparison sorting. Given an array $y$ having $m$ distinct integers, such that each integer is represented by at most $\kappa$ digits, the algorithm returns all the elements of $y$ sorted in ascending order. If two integers are identical, then their index in the array $y$ is used to determine their relative order, \ie for two identical integers $y[i] \equiv y[j]$, we assume that $y[i]$ precedes $y[j]$ if $i \leq j$ and $y[i]$ succeeds $y[j]$ otherwise. Since the presented suffix array construction algorithm runs on strings over $\Sigma = \mathbb{N} \cup \{-1\}$, then we can use the same algorithm, which we refer to as the \emph{parallel string sorting} algorithm, to sort an array of $m$ strings or tuples, each of fixed length $\kappa$. In this case, the algorithm has $O(\kappa \frac{m}{p})$ local computation and communication costs and requires $O(1)$ synchronisation costs. The algorithm, given an input array $y$ of $m$ strings over $\Sigma$, with each string of length at most $\kappa$, works as follows.

\vspace{4mm}

\noindent \textbf{Algorithm 2.} \textit{Parallel String Sorting}

\noindent \textbf{Parameters:} integer $m \geq p^3$; integer $\kappa$

\noindent \textbf{Input:} array of strings $y = y[0] \ldots y[m-1]$, with each string over $\Sigma = \mathbb{Z}$ and of size $\kappa$

\noindent \textbf{Output:} array $y$ ordered in ascending lexicographic order

\noindent \textbf{Description:}

The input array $y$ is assumed to be equally distributed among the $p$ processors, with every processor $\pi \in \frange{0}{p-2}$, assigned the elements $y\srange{\frac{m}{p} \pi}{\frac{m}{p} \left (\pi+1 \right )}$, and processor $p-1$ assigned elements $y\srange{\frac{m}{p} \left ( p-1 \right )}{m}$. Note that each processor holds $\frac{m}{p}$ elements, except the last processor $p-1$, which may hold fewer elements. We call this type of distribution of elements among the $p$ processors a \emph{block distribution}. Each processor $\pi$ first locally sorts sub-array $y_{\pi}$, using radix sorting, and then chooses $p + 1$ equally spaced samples from the sorted sub-array, including the minimum and maximum values of $y_{\pi}$. These samples, which we call \emph{primary samples}, are sent to processor $0$. Having received $(p+1)p$ primary samples, each of which is a string of length $\kappa$, processor $0$ locally sorts these samples, using radix sorting, and chooses $p + 1$ sub-samples, including the minimum and maximum values of the primary samples.  These chosen sub-samples, which we call \emph{secondary samples}, partition the elements of $y$ into $p$ blocks $Y_0, \ldots, Y_{p-1}$. The secondary samples are broadcast to every processor, and each processor $\pi$ then uses the secondary samples to partition its sub-array $y_{\pi}$ into the $p$ sub-blocks $Y_{0,\pi}, \ldots, Y_{p-1,\pi}$. Each processor $\pi$ collects the sub-blocks $Y_{\pi,\chi}$ from processors $\chi \in \srange{0}{p}$, \ie all the elements of $Y_{\pi}$, and locally sorts these elements using radix sorting. The array $y$ is now sorted in ascending lexicographic order, however, it might not be equally distributed among the processors, so an extra step is performed to ensure that each processor has $\frac{m}{p}$ elements of the sorted array. Note that each primary and secondary sample also has the index of the sample in $y$ attached to it, so that any ties can be broken. \Square

The parallel suffix array construction algorithm presented below requires that the input string $x$ of size $n$ be distributed equally among the $p$ processors, using a block distribution, prior to the algorithm being called. Therefore, each processor $\pi \in \frange{0}{p-2}$ initially holds the elements $x\srange{\frac{n}{p} \pi}{\frac{n}{p} \left ( \pi+1 \right )}$, while processor $p-1$ holds elements $x\srange{\frac{n}{p} \left ( p-1 \right )}{n}$. We denote by $I_{\pi}$ the subset of the index set $\srange{0}{n}$ that indexes $x_{\pi}$, $\pi \in \srange{0}{p}$, \ie $I_{\pi} = \srange{\frac{n}{p} \pi}{\frac{n}{p} \left ( \pi + 1 \right )}$, for $\pi \in \frange{0}{p-2}$, and $I_{p-1} = \srange{\frac{n}{p} \left ( p-1 \right )}{n}$. We require that every processor $\pi \in \frange{0}{p-2}$, also holds a copy of the first $v-1$ characters of the substring $x_{\pi+1}$, where $v$ is a parameter of the algorithm, to be able to locally construct its subset of super-characters. Finally, we we use the same indexing for $a$ and $a_{\pi}$, \ie $a[i] = a_{\pi}[i]$.

The algorithm is initially called on string $x$ of length $n$, with parameters $n$ and $v = 3$.

\vspace{4mm}

\noindent \textbf{Algorithm 3.} \textit{Parallel Suffix Array Construction}

\noindent \textbf{Parameters:} integer $n \geq p^{\frac{9}{2}}$; integer $v \in \frange{3}{n}$

\noindent \textbf{Input:} string $x = x[0] \ldots x[n-1]$ over alphabet $\Sigma = \srange{0}{n}$

\noindent \textbf{Output:} suffix array $SA_x = SA_x[0] \ldots SA_x[n-1]$

\noindent \textbf{Description:}

\noindent \textit{Recursion base}

Recall that if all the characters of $x$ are distinct, then $SA_x$ can be obtained by sorting the characters of $x$ in ascending order. Therefore, we call Algorithm 2 on string $x$ with parameters $m = n$ and $\kappa = 1$. When the algorithm returns with the sorted list of characters, which we call $x'$, each processor $\pi$, holds the sub-list $x'_{\pi}$ of size $\frac{n}{p}$, and checks for character uniqueness in its sub-list. If all the characters of each sub-list are distinct, then, each processor $\pi \in \frange{0}{p-2}$, checks with its neighbour $\pi + 1$ to ensure that $x'[\frac{n}{p}(\pi+1)-1] \not = x'[\frac{n}{p}(\pi+1)]$. If every character is distinct then each character in the sorted list $x'$ is replaced by its index in $x$ and $x'$ is returned. However, if at any point in this process two identical characters are found, then the following steps are performed:

\noindent \textit{Step 0 - Sample construction and initialisation}

Every processor $\pi$, constructs the difference cover $D$ of $\mathbb{Z}_v$ as discussed in Section \ref{sec:DiffCov}. Then, each processor $\pi$, for each $k \in \srange{0}{v}$, defines the subset ${B_k}_{\pi} = \{i \in I_{\pi} \mid i \bmod v=k\}$. This partitions each set of indices $B_k$ into $p$ subsets of size about $\frac{n}{pv}$. The subset $C_{\pi}$ of the difference cover sample $C$ is then constructed by every processor $\pi$, such that $C_{\pi} = \cup_{k \in D} {B_k}_{\pi}$. We denote by ${S_k}_{\pi}$, $k \in \srange{0}{v}$ and $\pi \in \srange{0}{p}$, the set of suffixes $s_i$, $i \in {B_k}_{\pi}$.

Finally, every processor $\pi$ also declares the array $rank_{\pi}$, of size $\frac{n}{p} + v$ for $\pi \in \frange{0}{p-2}$, and size $n - \frac{n}{p} (p-1) + v$ for $\pi = p-1$. Each element of $rank_{\pi}$ is initialised by -1. Note that the size of each $rank_{\pi}$, $\pi \in \frange{0}{p-2}$, follows from the fact that each such processor requires a copy of the first $v$ elements of $rank_{\pi+1}$ in order to be able to locally construct the tuples associated with all the non-sample characters in $x_{\pi}$.

\noindent \textit{Step 1 - Sort sample suffixes}

For every processor $\pi$, we define, for each $k \in D$, the substring of super-characters ${X_k}_{\pi} = \bigodot_{i \in {B_k}_{\pi}} \overline{x\srange{i}{i+v}}$, such that $|{X_k}_{\pi}| = \frac{n}{pv}$. Note that every substring $x\srange{i}{i+v}$ is locally available for all $i \in C_{\pi}$, due to the padding convention and the distribution of $x$ among the processors. Then, construct the string of super-characters $X$, as discussed in Section $\ref{sec:SeqAlg}$. This string is distributed among the $p$ processors, with each processor having $|D| \frac{n}{pv}$ super-characters. Note that it is not necessary to actually construct $X$, since the position of each ${X_k}_{\pi}$, and, therefore, the index of each super-character $\overline{x\srange{i}{i+v}}$, $i \in C$, in $X$ can be calculated by every processor $\pi$. However, this is done for simplicity. Algorithm 2 is then called on string $X$ with parameters $m = |D|\frac{n}{v}$ and $\kappa = v$. After sorting, a rank is assigned to each super-character in its sorted order, with any identical super-characters given the same rank, and the string $X'$ is constructed, as discussed in Section \ref{sec:SeqAlg}. Note that $X'$ is already equally distributed among the processors.

The algorithm is then called recursively on the string $X'$ with parameters $n = |X'|$ and $v' = \min (v^{\sfrac{5}{4}}, |X'|)$, where $v'$ is the value of $v$ in the subsequent recursion level. If $|X'| \leq \frac{n}{p}$, then $X'$ is sent to processor $0$, which calls the sequential suffix array algorithm on $X'$ with parameters $n = |X'|$ and $v = 3$. A detailed discussion on the bound of $v' = \min (v^{\sfrac{5}{4}}, |X'|)$ and its impact on the synchronisation costs of the algorithm is given later in this section.

When the recursive call returns with $SA_{X'}$, the rank of each $s_i$ in $SA_{X'}$, $i \in {C_k}_{\pi}$, $\pi \in \srange{0}{p}$, is recorded in $rank_{\pi}$. Also, a copy of the first $v$ elements of $rank_{\pi}$, for $\pi \in \srange{1}{p}$, is kept in $rank_{\pi - 1}$. The order of each suffix $s_i$ within each set $S_k$, $k \in D$, is stored by each processor $\pi$, for $i \in I_{\pi}$.

\noindent \textit{Step 2 - Find the order of the non-sample suffixes within each set $S_k$, $k \in \mathbb{Z}_v \setminus D$}

For each $k \in \mathbb{Z}_v \setminus D$, consider any $l_k \in \srange{1}{v}$ such that $(k + l_k) \bmod v \in D$. We define the tuple $t_i = (x[i], x[i+1], \ldots, x[i+l_k -1], rank[i+l_k])$, for each character $x[i]$, $i \in I_{\pi} \setminus C_{\pi}$, $\pi \in \srange{0}{p}$ and $k = i \mod v$. Note that every character in the tuple can be constructed locally on processor $\pi$.

Then, every processor $\pi \in \srange{0}{p}$ constructs the subsequence of tuples $(t_i)_{i \in {B_k}_{\pi}}$, for each subset ${B_k}_{\pi}$, $k \in \mathbb{Z}_v \setminus D$. Therefore, each sequence $(t_i)_{i \in {B_k}}$ is the concatenation of the subsequences $(t_i)_{i \in {B_k}_{\pi}}$ in ascending order of $\pi$. Recall from Section \ref{sec:SeqAlg}, that the number of sequences $(t_i)_{i \in {B_k}}$ to be sorted is $v - |D|$, and that each sequence contains $\frac{n}{v}$ tuples, of length at most $v$. Therefore, each processor holds about $\frac{n}{vp}$ tuples of each sequence.

Each sequence is then sorted using Algorithm 2 with parameters $m = \frac{n}{v}$ and $\kappa$ being the length of the tuples in the sequence, which is at most $v$. After each sequence is sorted, the order of each non-sample suffix $s_i$ within each set $S_k$, $k \in \mathbb{Z}_v \setminus D$, is stored by each processor $\pi$, $i \in I_{\pi}$.

\noindent \textit{Step 3 - Sort all suffixes by first $v$ characters}

Let each suffix $s_i$, $i \in \srange{0}{n}$, of $x$ be represented by the substrings $x\srange{i}{i+v}$. These substrings are sorted using Algorithm 2 with parameters $m = n$ and $\kappa = v$. The index of each substring in $x$ is used to determine the order of identical substrings. After sorting, the suffixes of $x$ will have been partitioned into the sets $S^{\alpha}$, $\alpha \in (\Sigma \cup \{-1\})^v$, as discussed in Section \ref{sec:SeqAlg}.

\noindent \textit{Step 4 - Merge and complete the suffix ordering}

Recall from Section \ref{sec:SeqAlg} that, each set $S^{\alpha}$, $\alpha \in (\Sigma \cup \{-1\})^v$, is partitioned into at most $v$ subsets $S^{\alpha}_{k}$, $k \in \srange{0}{v}$, and that the order of the suffixes within each such subset has been found in the previous steps. Ordering a set $S^{\alpha}$ is achieved through a $v$-way merging procedure based on Lemma \ref{lem:DiffCov}. For every two subsets $S^{\alpha}_{k'}$ and $S^{\alpha}_{k''}$, $k' , k'' \in \srange{0}{v}$, we choose any $l \in \srange{0}{v}$ such that $(k' + l) \bmod v$ and $(k'' + l) \bmod v$ are both in $D$. Then, comparing two suffixes $s_i \in S_{k'}$ and $s_j \in S_{k''}$ only requires the comparison of $rank[i + l]$ and $rank[j+l]$.

Therefore, in order to sort every element of $S^{\alpha}$ we require, for each element of $S^{\alpha}$, the order of the element within the subset $S^{\alpha}_{k}$ it belongs to and at most $|D|$ values from the array $rank$. Hence, at most $(|D| + 1) \frac{n}{p}$ values need to be received by each processor. Note that the order of each suffix $s_i$, $i \in \srange{0}{n}$, within the set $S_k$, $i \bmod v = k$, is stored on processor $\pi$, $i \in I_{\pi}$, as is $rank[i + l]$, for any $l \in \srange{0}{v}$.

After the sorting procedure in the previous step, the suffixes of a set $S^{\alpha}$, $\alpha \in \Sigma^v$, are contiguous and can be either contained within a single processor, or span two or more processors. If $S^{\alpha}$ is contained within one processor, then this processor locally merges each of the subsets of $S^{\alpha}$. If the set spans two processors $\pi', \pi'' \in \srange{0}{p}$, then, for each of the suffixes $s_i \in S^{\alpha}$, $i \in \srange{0}{n}$, on processor $\pi''$, the values required to merge $s_i$ into the ordered $S^{\alpha}$ are sent to $\pi'$. Processor $\pi'$ then locally merges each of the subsets of $S^{\alpha}$. Otherwise, if the set $S^{\alpha}$ spans more than two processors, the following procedure is performed.

Let $p'$ be the number of processors that the set $S^{\alpha}$ spans. Then, $S^{\alpha}$ is equally divided among the $p'$ processors, such that each is assigned $\frac{|S^{\alpha}|}{p'}$ elements. Again, note that the actual suffixes $s_i \in S^{\alpha}$, $i \in \srange{0}{n}$, are not communicated, but only the values required by the merging process are, \ie at most $|D| + 1$ values for each suffix in $S^{\alpha}$.

Each of the $p'$ processors locally sorts its assigned elements of $S^{\alpha}$, using the $v$-way merging procedure, and chooses $p' + 1$ equally spaced primary samples from the sorted elements, including the minimum and maximum elements. Every primary sample is sent to one of the $p'$ processors that is chosen as the designated processor. Therefore, this designated processor receives $(p' + 1) p'$ primary samples, which it sorts locally using the $v$-way merging procedure. It then chooses $p' + 1$ equally spaced secondary samples from the merged primary samples, including the minimum and maximum primary samples, that partition $S^{\alpha}$ into $p'$ blocks. These secondary samples are broadcast to the $p'$ processors such that each processor can partition its assigned elements into $p'$ sub-blocks. Every processor then collects all the sub-blocks that make up a unique block and locally merges the received elements.

Note that a processor can only have elements from at most two sets that span across three or more processors. Therefore, this procedure can be done in parallel for each set $S^{\alpha}$. After all the sets $S^{\alpha}$ have been sorted, all the suffixes of $x$ have been ordered and the suffix array is returned. \Square

\subsection{Algorithmic Analysis}

The presented suffix array construction algorithms are recursive, and the number of levels of recursion required for the algorithms to terminate depends on the factor by which the size of the input string is reduced in successive recursive calls. While the number of levels of recursion does not influence the running time of the sequential algorithm, in BSP this determines the synchronisation costs of the algorithm, and, therefore, we want to reduce it to a minimum. Before detailing the costs of each step of the algorithm we explain how changing the sample size at each subsequent level of recursion results in $O(\log \log p)$ levels of recursion.

We refer to each level of recursion of the algorithm as round $i$, $i \geq 0$. Then, we denote by $n_i$, $v_i$ and $D_i$ the size of the input string, the parameter $v$ and the difference cover $D$ of $\mathbb{Z}_{v_i}$, respectively, in round $i$.

Recall from Section \ref{sec:DiffCov} that, the maximum size of a difference cover $D$ of $\mathbb{Z}_v$, for any positive integer $v$, that can be found in time $O(\sqrt{v})$ is $\sqrt{1.5 v} + 6$, \ie $|D| = O(v^{\sfrac{1}{2}})$. Therefore, for the sake of simplicity, in our cost analysis we assume that $|D_i| = {v_i}^{\sfrac{1}{2}}$.

The analysis given in Table \ref{tab:AlgAnal} shows how changing the sampling rate affects the parameters $v$ and $n$ in subsequent recursive calls. Recall from Section \ref{sec:SeqAlg} that, the cost of each level of recursion in the sequential algorithm is $O(v_i n_i)$. Therefore, the table also shows that the order of work done decreases in subsequent recursive calls.

\begin{table}[t]
\centering
\caption{Algorithm analysis}
\begin{tabular}{|c|c|c|c|c|}
\hline
Round $i$ & $v_i$ & $|D_i|$ & $n_i$ & Work\\
\hline
$0$ & $v$ & $O\left ( v^{\frac{1}{2}} \right )$ & $n$ & $O\left ( v \cdot n \right )$\\
\hline
$1$ & $v^{(\frac{5}{4})}$ & $O\left ( (v^{(\frac{5}{4})})^{\frac{1}{2}} \right )$ & $\frac{v^{\frac{1}{2}}}{v} \cdot n = v^{-\frac{1}{2}} \cdot n$ & $O\left ( v^{(\frac{5}{4})} \cdot v^{(-\frac{1}{2})} \cdot n \right )$\\
\hline
$2$ & $v^{(\frac{5}{4})^2}$ & $O\left ( (v^{(\frac{5}{4})^2})^{\frac{1}{2}} \right )$ & $\frac{(v^{(\frac{5}{4})})^{\frac{1}{2}}}{v^{(\frac{5}{4})}} \cdot v^{-\frac{1}{2}} \cdot n = v^{(-\frac{9}{8})} \cdot n$ & $O\left ( v^{(\frac{5}{4})^2} \cdot v^{(-\frac{9}{8})} \cdot n \right )$\\
\hline
$i$ & $v^{(\frac{5}{4})^i}$ & $O\left ( v^{(\frac{5}{4})^i(\frac{1}{2})} \right )$ & $v^{-2(\frac{5}{4})^i + 2} \cdot n$ & $O\left ( v^{-(\frac{5}{4})^i + 2} \cdot n \right )$\\
\hline
$\log_{\frac{5}{4}} \left ( \log_v p^{\frac{1}{2}} + 1 \right )$ & $v \cdot p^{\frac{1}{2}}$ & $O\left ( v^{\frac{1}{2}} \cdot p^{\frac{1}{4}} \right )$ & $v^{2} \cdot \frac{n}{p}$ & $O\left ( v^{3} \cdot \frac{n}{p^{\frac{1}{2}}} \right )$\\
\hline
\end{tabular}
\label{tab:AlgAnal}
\end{table}

The results in Table \ref{tab:AlgAnal} clearly show that, if the algorithm is initially called on a string of size $n$, with parameter $v = 3$, on a $BSP(p,g,l)$ machine, then the size of the input converges towards $\frac{n}{p}$ super-exponentially. In fact,  after $\log_{\sfrac{5}{4}} ( \log_{3} p^{\sfrac{1}{2}} + 1 ) = O(\log \log p)$ levels of recursion, the size of the input string is $O(\frac{n}{p})$, and in the subsequent level of recursion the suffix array is computed sequentially on processor $0$. Note that the value $\frac{5}{4}$ as a power of $v$ is not the only one possible. In fact, any value $1 < a < \frac{3}{2}$ can be used, but $a = \frac{5}{4}$ is used for simplicity. Finally, note that $v_i > n_i$ only after $O(\log \log p)$ levels of recursion, at which point the algorithm is called sequentially on a single processor.

Having determined the number of recursive calls required by the algorithm, the cost of each step of the algorithm is analysed. For each step, the costs of the first round of the algorithm are specified below, along with the costs of round $\log_{\sfrac{5}{4}} ( \log_3 p^{\sfrac{1}{2}} + 1 )$, which we call the \emph{critical round}, since this is the round immediately before the algorithm is called sequentially on a string of length less than $\frac{n}{p}$.

In the recursion base, the costs are dominated by those of Algorithm 2, \ie $O(\frac{n_i}{p})$ local computation and communication cost. Therefore, in the first round the local computation and communication costs are $O(\frac{n}{p})$, and in the critical round these costs are $O(\frac{n}{p^2})$. A constant number of synchronisation steps is required.

In step 0, constructing the difference cover $D_i$ has running time $O(\sqrt{v_i})$, \ie $O(1)$ in the first round and $O(p^{\sfrac{1}{4}})$ in the critical round. Constructing the subsets $C_{\pi}$, independently for each processor $\pi \in \srange{0}{p}$, has $O(|D_i| \frac{n_i}{pv_i})$ local computation cost, \ie $O(\frac{n}{p})$ in the first round and $O(\frac{n}{p^{\sfrac{9}{4}}})$ in the critical round. Finally, declaring and initialising $rank_{\pi}$ requires $O(\frac{n_i}{p} + v_i)$ work, \ie $O(\frac{n}{p})$ in the first round and $O(\frac{n}{p^2})$ in the critical round. A single synchronisation step is required, with no communication.

In step 1, the costs are dominated by the construction of the string of super-characters $X$ and the call to Algorithm 2, leading to $O(|D_i|\frac{n_i}{p})$ local computation and communication costs. Therefore, the costs of this step in the first round are $O(\frac{n}{p})$ local computation and communication, while in the critical round these costs are $O(\frac{n}{p^{\sfrac{7}{4}}})$. A constant number of synchronisation steps is required in each round.

In step 2, the costs are again dominated by the call to Algorithm 2 for each sequence of tuples. The size of each sequence is $\frac{n_i}{v_i}$, and the size of each tuple is at most $v_i$. Therefore, the local computation and communication costs to sort each sequence are $O(\frac{n_i}{p})$, \ie $O(\frac{n}{p})$ in the first round and $O(\frac{n}{p})$ in the critical round. Each sequence is sorted independently using Algorithm 2, and, since the number of sequences to be sorted, $v_i - |D_i|$, is always less than $p$, then each sequence can be sorted in parallel in each round by having a different designated processor for each call to Algorithm 2. Therefore, the number of synchronisation steps required is always constant. Recall that Algorithm 2 requires slackness, $m \geq p^{3}$. Since, in the critical round, Algorithm 2 is called on a sequence of length $9 \frac{n}{p}$, we require that $n \geq p^{\sfrac{9}{2}}$. Note that this slackness can be reduced by sorting sequences locally if each sequence fits on a separate processor, however, such detail is beyond the scope of this paper and will be given in a journal version of this paper.

The cost of step 3 is simply the cost of Algorithm 2 on a string of size $n_i$ with $\kappa = v_i$, \ie $O(v_i \frac{n_i}{p})$ local computation and communication costs and $O(1)$ synchronisation steps. Therefore, in the first round the local computation and communication costs are $O(\frac{n}{p})$ and these costs in the critical round are $O(\frac{n}{p^{\sfrac{3}{2}}})$.

In step 4, obtaining, for each suffix of $x$, the information required for the sorting each set $S^{\alpha}$ using a $v$-way merging procedure has $O(|D_i|\frac{n_i}{p})$ local computation and communication costs, \ie $O(\frac{n}{p})$ in the first round and $O(\frac{n}{p^{\sfrac{3}{2}}})$ in the critical round. Then, sorting a set $S^{\alpha}$ that is contained on a single processor has $O(|S^{\alpha}| v_i)$ local computation costs, and no communication is required. Note that in this case $|S^{\alpha}| < \frac{n_i}{p}$, so the local computation costs are $O(\frac{n}{p})$ in the first round and $O(\frac{n}{p^{\sfrac{3}{2}}})$ in the critical round. If $S^{\alpha}$ spans two processors, then we send all the elements of the set to one of the two processors. Therefore, since each processor has $\frac{n_i}{p}$ suffixes, $|S^{\alpha}| < 2 \frac{n_i}{p}$, so the costs of sorting this set are $O(v_i \frac{n_i}{p})$ local computation, $O(|D_i| \frac{n_i}{p})$ communication and $O(1)$ synchronisation steps; \ie $O(\frac{n}{p})$ local computation and communication costs in the first round and $O(\frac{n}{p^{\sfrac{3}{2}}})$ in the critical round.

Finally, if a set $S^{\alpha}$ spans $p' > 2$ processors, then $|S^{\alpha}| > (p' - 1) \frac{n_i}{p}$. In this case a procedure similar to the parallel radix sorting algorithm on $p'$ processors is performed. In fact, the only difference between the two is that $v$-way merging is used, instead of radix sorting, to locally sort the suffixes on each of the $p'$ processors. Since the $v$-way merging procedure on $n$ elements has the same asymptotic costs as the radix sorting procedure on an array of $n$ strings each of size $v$, over an alphabet $\Sigma = \mathbb{Z}$, then the local computation cost for this procedure is $O(v_i \frac{n_i}{p})$ and the communication cost is $O(|D_i| \frac{n_i}{p})$. Therefore, both these costs are $O(\frac{n}{p})$ in the first round and $O(\frac{n}{p^{\sfrac{3}{2}}})$ in the critical round. Since each such set can be merged independently in parallel, then a constant number synchronisation steps is required.

Note that, in the $i^{th}$ level of recursion, each step has local computation cost $O(v_i \frac{n_i}{p})$, communication cost $O(v_i \frac{n_i}{p})$ and $O(1)$ synchronisation costs. Also note that, as shown in table \ref{tab:AlgAnal}, $O(v_i n_i)$ decreases super-exponentially in each successive  level of recursion, and, therefore, the order of work done in each round of the BSP algorithm also decreases super-exponentially. Since the presented parallel suffix array construction algorithm is initially called on a string of size $n$ with parameter $v = 3$, the algorithm has $O(\frac{n}{p})$ local computation and communication costs and requires $O(\log \log p)$ synchronisation steps.

%% file: Conclusion.tex
\section{Conclusion}
\label{sec:Conc}

In this paper we have presented a deterministic BSP algorithm for the construction of the suffix array of a given string. The algorithm runs in optimal $O(\frac{n}{p})$ local computation and communication, and requires a near optimal $O(\log \log p)$ synchronisation steps.

The method of regular sampling in coarse-grained algorithms has been used to solve the sorting \cite{ShiSchaeffer,ChanDehne}, and 2D and 3D convex hulls \cite{Tiskin_convexhull} problems. Random sampling has been used to solve the maximal matching problem and provide an approximation to the minimum cut problem \cite{LattanziMoseleySuriVassilvitskii} in a parallel context. An extension of the regular sampling technique, which we call \textit{accelerated sampling}, was introduced by Tiskin \cite{Tiskin} to improve the synchronisation upper bound of the BSP algorithm for the selection problem. The same technique was used here to improve the synchronisation upper bounds of the suffix array construction problem. Accelerated sampling is a theoretically interesting technique, allowing, in specific cases, for an exponential factor improvement in the number of synchronisation steps over existing algorithms.

It is still an open question as to whether the synchronisation cost of the suffix array construction problem and the selection problem can be reduced to the optimal $O(1)$ while still having optimal local computation and communication costs. Another open question is whether further applications of the sampling technique, whether regular, random or accelerated, are possible.